\documentclass[journal]{IEEEtran}

\usepackage{pgfplots}
\usepackage[bookmarks,colorlinks]{hyperref}
\usepackage[linesnumbered,ruled,lined]{algorithm2e}
\usepackage{enumitem}
\usepackage{algpseudocode}
\usetikzlibrary{shapes.multipart,intersections}
\usepackage{cite}
\usepackage{amsmath,amssymb,amsfonts,amsthm,steinmetz}
\usepackage{graphicx}
\usepackage{mathrsfs}  
\usepackage{textcomp}
\usepackage{acronym}
\usepackage{xcolor}
\usepackage{upgreek,xspace}
\usepackage{array}
\usepackage{tikz}
\usetikzlibrary{calc}
\makeatletter
\newcommand{\gettikzxy}[3]{%
  \tikz@scan@one@point\pgfutil@firstofone#1\relax
  \edef#2{\the\pgf@x}%
  \edef#3{\the\pgf@y}%
}
\makeatother

\usepackage[draft]{todonotes}

\usepackage{esvect}

\usepackage{times}
\usepackage{bm}
\usepackage{amsmath}
\usepackage{amssymb}
\usepackage{stmaryrd}
\usepackage{babel}
\usepackage{graphics, graphicx}
\usepackage{xcolor}
\usepackage{gensymb}
\usepackage{cite}
\usepackage{enumitem}
\usepackage{url}

\begin{document}

\title{Beyond-Diagonal RIS Prototype \\and Performance Evaluation}

\author{Jean~Tapie,~Matteo~Nerini,~\IEEEmembership{Member,~IEEE},~Bruno~Clerckx,~\IEEEmembership{Fellow,~IEEE},~and~Philipp~del~Hougne,~\IEEEmembership{Member,~IEEE}
\thanks{This work was supported in part by the ANR France 2030 program (project ANR-22-PEFT-0005), the ANR PRCI program (project ANR-22-CE93-0010), the French Defense Innovation Agency (project 2024600), the French Region of Brittany (project TUNSY), the European Union's European Regional Development Fund, and the French Region of Brittany and Rennes Métropole through the contrats de plan État-Région program (projects ``SOPHIE/STIC \& Ondes'' and ``CyMoCoD'').}
\thanks{
J.~Tapie and P.~del~Hougne are with Univ Rennes, CNRS, IETR - UMR 6164, F-35000, Rennes, France (e-mail: \{jean.tapie; philipp.del-hougne\}@univ-rennes.fr).
}
\thanks{
M.~Nerini and B.~Clerckx are with the Department of
Electrical and Electronic Engineering, Imperial College London, SW7 2AZ
London, U.K (e-mail: \{m.nerini20; b.clerckx\}@imperial.ac.uk).
}
\thanks{\textit{(Corresponding Author: Philipp del Hougne.)}}
}

\maketitle

\begin{abstract}
We present the first experimental prototype of a reflective beyond-diagonal reconfigurable intelligent surface (BD-RIS), i.e., a RIS with reconfigurable inter-element connections. Our BD-RIS consists of an antenna array whose ports are terminated by a tunable load network. The latter can terminate each antenna port with three distinct individual loads or connect it to an adjacent antenna port. Extensive performance evaluations in a rich-scattering environment validate that inter-element connections are beneficial. Moreover, we observe that our tunable load network's mentioned hardware constraints significantly influence, \textit{first}, the achievable performance, \textit{second}, the benefits of having inter-element connections, and, \textit{third}, the importance of mutual-coupling awareness during optimization. 
\end{abstract}


\section{Introduction}

Tailoring wireless channels with reconfigurable intelligent surfaces (RISs) is emerging as an enabling technology of future wireless networks. An RIS is an array of elements with tunable scattering properties. A typical reflective RIS can be characterized as an array of antenna elements whose ports are terminated by tunable loads. Placed inside a radio environment, the load configuration of an RIS hence parametrizes the wireless channels and enables their optimization. To further increase the control, beyond-diagonal RIS (BD-RIS) with tunable inter-element connections have recently been proposed~\cite{shen2021modeling}. The RIS element ports of a BD-RIS are terminated by a tunable load network whose scattering matrix is not restricted to being diagonal.

Theoretical studies on BD-RIS increasingly seek to consider conditions that match the physical reality, e.g., by accounting for mutual coupling (MC) effects between the RIS elements~\cite{li2024beyond}. However, it is still commonly assumed that the tunable load network can realize any arbitrary scattering matrix that is unitary and reciprocal. 
Moreover, the influence of scattering in the radio environment (including structural scattering of the RIS elements~\cite{king1949measurement}) on the MC between RIS elements and the end-to-end wireless channels is typically simplified or neglected. Recently, the first experimentally grounded study on BD-RIS was presented in~\cite{del2025physics}, based on experimental channel measurements in a rich-scattering environment and assuming a group-connected load network parametrized by realistic 1-bit tunable lumped elements (PIN diodes).
Nonetheless, to date no full experimental realization of a reflective BD-RIS prototype has been reported.\footnote{During the preparation of this Letter, a preprint presenting a simultaneously transmitting and reflecting RIS (STAR-RIS) with tunable couplings linking each front element to its back counterpart appeared~\cite{ming2025hybrid}. Although this STAR-RIS can be characterized as BD-RIS, its inter-element connections are indispensable for the STAR-RIS transmission mode, precluding an analysis that quantifies benefits of inter-element connections.} There is hence an urgent need to  fill the current research gap on experimental realizations of BD-RIS prototypes.

Besides this BD-RIS context, reconfigurable inter-element connections also play an important role in lifting ambiguities in backscatter channel measurements~\cite{denicke2012application} and, more generally, the ``Virtual Vector Network Analyzer'' (Virtual VNA) concept~\cite{del2024virtual2p0,del2025virtual3p0}. A VNA is a standard microwave engineering instrument that measures a multi-port network scattering matrix by injecting and receiving waves through \textit{all} system ports.
The Virtual VNA unambiguously estimates the full scattering matrix of a multi-port system by injecting and receiving waves \textit{only via a fixed subset} of ports while terminating the remaining ``not-directly-accessible'' (NDA) ports with a tunable load network~\cite{del2024virtual2p0,del2025virtual3p0}. The latter must be able to switch the termination of the NDA system ports between at least three distinct individual loads and connections to adjacent ports~\cite{del2024virtual2p0,del2025virtual3p0}. A printed-circuit-board (PCB) realization of such a tunable load network was recently presented in~\cite{tapie2025scalable}. 
Interestingly, a related circuit topology has been theoretically proposed for implementing a specific BD-RIS architecture, denoted as tridiagonal RIS~\cite{matteo_graph}.
The tridiagonal RIS is a specific instance of tree-connected BD-RISs, which are the BD-RIS architectures theoretically proved to achieve maximum performance in single-user systems with the minimum number of tunable loads~\cite{matteo_graph}.
Hence, we re-purpose the Virtual VNA PCB from~\cite{tapie2025scalable} in this Letter to build the first BD-RIS prototype by connecting the ports of an antenna array to this PCB.

Our contributions are summarized as follows. \textit{First}, we present the first experimental prototype of a reflective BD-RIS. \textit{Second}, we leverage the Virtual VNA technique to unambiguously estimate all parameters of a physics-consistent multi-port network theory (MNT) model of our experimental BD-RIS-parametrized rich-scattering environment.
\textit{Third}, we report the results of extensive performance evaluations, benchmarking BD-RIS against conventional RIS, benchmarking our hardware-constrained system against an idealized one, and assessing the importance of MC-awareness during optimization.

\textit{Notation:} $\mathbf{A}_\mathcal{BC}$ denotes the block of the matrix $\mathbf{A}$ whose row [column] indices are in the set $\mathcal{B}$ [$\mathcal{C}$]. $\|\mathbf{A}\|$ denotes the spectral norm of $\mathbf{A}$. $\mathbf{A}^\top$ denotes the transpose of $\mathbf{A}$. $\mathbf{A}^H$ denotes the conjugate transpose of $\mathbf{A}$. $\mathbf{I}_a$ denotes the $a \times a$ identity matrix. 

\section{System Model}

We physics-consistently describe the static components of a radio environment comprising $N_\mathrm{T}$ transmitters, $N_\mathrm{R}$ receivers, and $N_\mathrm{S}$ RIS elements as a passive, linear, time-invariant $N$-port network characterized by its scattering matrix $\mathbf{S}\in\mathbb{C}^{N\times N}$, where $N=N_\mathrm{T}+N_\mathrm{R}+N_\mathrm{S}$. $\mathbf{S}$ compactly accounts for scattering objects (including structural scattering at the antennas and RIS elements) without requiring their explicit description~\cite{tapie2023systematic}. 
We use the same reference impedance $Z_0$ at all ports to define $\mathbf{S}$ and assume for simplicity that the signal generators and detectors (attached to the transmitting and receiving antennas, respectively) are matched to $Z_0$. We further describe the tunable components of a BD-RIS-parametrized radio environment in terms of the scattering matrix $\mathbf{S}_\mathrm{L}\in\mathbb{C}^{N_\mathrm{S}\times N_\mathrm{S}}$ that characterizes the tunable load network. 
Both $\mathbf{S}$ and $\mathbf{S}_\mathrm{L}$ are symmetric (due to reciprocity) and their singular values do not exceed unity (due to passivity). We detail in Sec.~\ref{sec_experiment} how we estimate $\mathbf{S}$ using the Virtual VNA technique and what realizations of $\mathbf{S}_\mathrm{L}$ are feasible based on the measured scattering characteristics of our PCB. 

According to MNT~\cite{li2024beyond,del2025physics}, the end-to-end wireless channel $\mathbf{H}\in\mathbb{C}^{N_\mathrm{R}\times N_\mathrm{T}}$ is related to $\mathbf{S}$ and $\mathbf{S}_\mathrm{L}$ as follows:
\begin{equation}
    \mathbf{H} = {\mathbf{S}}_\mathcal{RT} +{\mathbf{S}}_\mathcal{RS} \left(\mathbf{S}_\mathrm{L}^{-1} - {\mathbf{S}}_\mathcal{SS}  \right)^{-1} {\mathbf{S}}_\mathcal{ST},
    \label{eq1}
\end{equation}
where $\mathcal{R}$, $\mathcal{T}$, and $\mathcal{S}$ denote the sets of port indices associated with receiving antennas, transmitting antennas, and RIS elements, respectively.\footnote{The matrices $\mathbf{S}_{\mathcal{RT}}\in\mathbb{C}^{N_\mathrm{R}\times N_\mathrm{T}}$, $\mathbf{S}_{\mathcal{RS}}\in\mathbb{C}^{N_\mathrm{R}\times N_\mathrm{S}}$, $\mathbf{S}_{\mathcal{ST}}\in\mathbb{C}^{N_\mathrm{S}\times N_\mathrm{T}}$, and $\mathbf{S}_{\mathcal{SS}}\in\mathbb{C}^{N_\mathrm{S}\times N_\mathrm{S}}$ are hence submatrices of $\mathbf{S}$ according to the partition
$\mathbf{S}=[
[\mathbf{S}_{\mathcal{TT}},\mathbf{S}_{\mathcal{TS}},\mathbf{S}_{\mathcal{TR}}]^\top,
[\mathbf{S}_{\mathcal{ST}},\mathbf{S}_{\mathcal{SS}},\mathbf{S}_{\mathcal{SR}}]^\top,
[\mathbf{S}_{\mathcal{RT}},\mathbf{S}_{\mathcal{RS}},\mathbf{S}_{\mathcal{RR}}]^\top]^\top$.} $ {\mathbf{S}}_\mathcal{SS} $ captures the MC between the RIS elements; with all entries of $ {\mathbf{S}}_\mathcal{SS} $ set to zero, (\ref{eq1}) specializes to the widespread MC-unaware simplified cascaded model:
\begin{equation}
    \mathbf{H}_\mathrm{casc} = {\mathbf{S}}_\mathcal{RT} +{\mathbf{S}}_\mathcal{RS} \mathbf{S}_\mathrm{L} {\mathbf{S}}_\mathcal{ST}.
    \label{eq2}
\end{equation}
Unless otherwise stated, we do \textit{not} use any simplification of (\ref{eq1}) such as neglecting MC as in (\ref{eq2}).

\section{Experimental Setup and Procedure}
\label{sec_experiment}

\begin{figure}
\centering
\includegraphics[width=1\columnwidth]{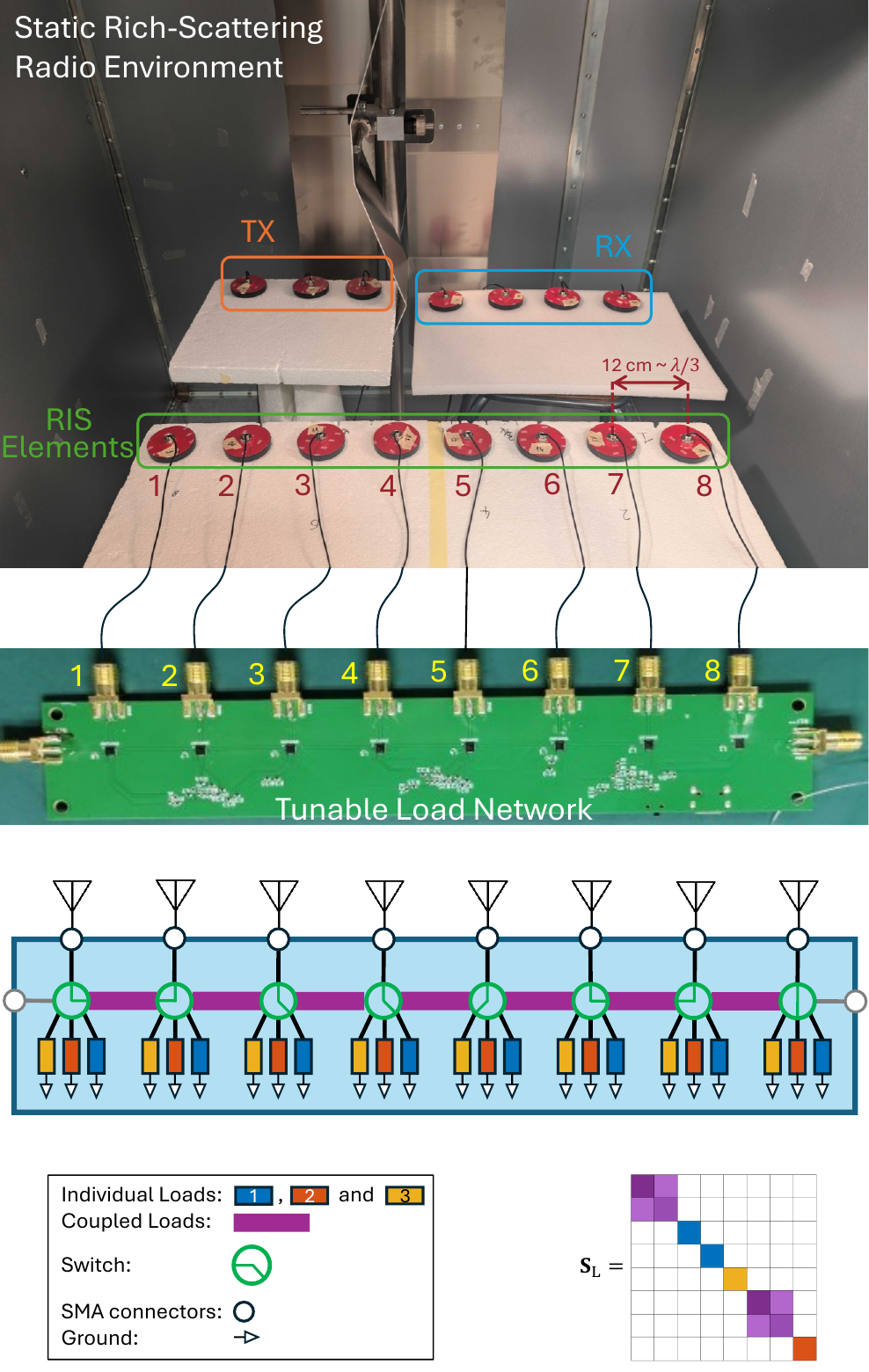}
\caption{Experimental BD-RIS-parametrized rich-scattering radio environment. Bottom: Schematic of the BD-RIS and illustration of the $\mathbf{S}_\mathrm{L}$ realization corresponding to the switch setting in the schematic. }
\label{fig1}
\end{figure}

As mentioned earlier, we built our BD-RIS by connecting (via SubMiniature version A connectors) a set of antennas serving as RIS elements to a PCB-realized tunable load network. In contrast to an integrated design, this modular design allows us to directly measure the available load characteristics, which is important for the application of the MNT model. Our experimentally realized BD-RIS-parametrized rich-scattering radio environment shown in Fig.~\ref{fig1} comprises $N=15$ commercial antennas (AEACBK081014-S698) placed inside a reverberation chamber (1.75~m~$\times$~1.50~m~$\times$~2.00~m). The reverberation chamber (including its mode stirrer) is static throughout all experiments. A paddle of the metallic mode stirrer blocks the line-of-sight between the $N_\mathrm{T}=3$ transmitting and the $N_\mathrm{R}=4$ receiving antennas. The remaining $N_\mathrm{S}=8$ antennas act as RIS elements and are connected to our PCB-realized tunable load network. The spacing between antennas (for TX and RX) or RIS elements is 12~cm, which is roughly a third of the wavelength in the consider 700-900~MHz range. 

Each of the PCB's eight ports (indicated by black circles in the schematic in Fig.~\ref{fig1}) feeds into a single-pole, five-throw (SP5T) radio-frequency switch (SKY13414-485LF), which can select among five different terminations. Three of those terminations are discrete loads (two strongly reflective with roughly opposite phases, one strongly absorptive), and the other two form connections to the switches on either side (i.e., coupled loads). The absorptive individual load is based on the integrated absorbing termination of the switch; the two reflective loads are realized with two 2.55~mm long delay lines, one of which is terminated by a 4~nH inductor. 

Our BD-RIS is hence a tridiagonal RIS~\cite{matteo_graph}, i.e., capable of populating the first super-diagonal and sub-diagonal of $\mathbf{S}_\mathrm{L}$. Due to our hardware constraints, we cannot populate two adjacent super-diagonal (and sub-diagonal) entries because a given switch cannot simultaneously connect a given RIS element to both adjacent RIS elements. 
Interestingly, this hardware constraint implies that not only $\mathbf{S}_\mathrm{L}$ but also the corresponding impedance matrix $\mathbf{Z}_\mathrm{L}$ and admittance matrix $\mathbf{Y}_\mathrm{L}$ are tri-diagonal.
Moreover, we cannot control the super-diagonal (and sub-diagonal) entries independently of the two closest diagonal entries. 
A schematic illustration of a realizable $\mathbf{S}_\mathrm{L}$ given our hardware constraints is shown on the bottom of Fig.~\ref{fig1}.\footnote{The schematic of our BD-RIS prototype shown in Fig.~\ref{fig1} resembles [Fig.~3,~\cite{matteo_graph}] wherein a tridiagonal RIS was theoretically proposed. Both architectures have in common that the realizable $\mathbf{Y}_\mathrm{L}$ are tridiagonal. 
Yet, a key difference lies in the tunability mechanism. [Fig.~3,~\cite{matteo_graph}] presents a circuit topology wherein a set of tunable lumped elements is connected (without delay or loss). Our prototype relies on a circuit topology with multiple static loads that are switched in or out of the signal path. The tunability hence resides within the switch rather than the loads. Each switch is itself a tunable six-port network. The latter can be separated into its static parts and a set of five tunable lumped elements. Therefore, based on the arguments put forth in~\cite{del2025physics}, ultimately a physics-compliant \textit{diagonal} representation exists for both the architecture in [Fig.~3,~\cite{matteo_graph}] and our prototype, however, with different numbers of tunable lumped elements. The diagonal representation implies that existing physics-compliant D-RIS algorithms can be directly applied to these BD-RIS architectures~\cite{del2025physics}.}

To configure our tunable load network, we send a 24-bit control word from Python over USB-serial to an STM32F103C8T6 microcontroller. The latter splits this into three 8-bit segments and clocks them into three 8-bit shift registers, which present the 3-bit digital DC control signals for each switch.

\begin{figure}
\centering
\includegraphics[width=1\columnwidth]{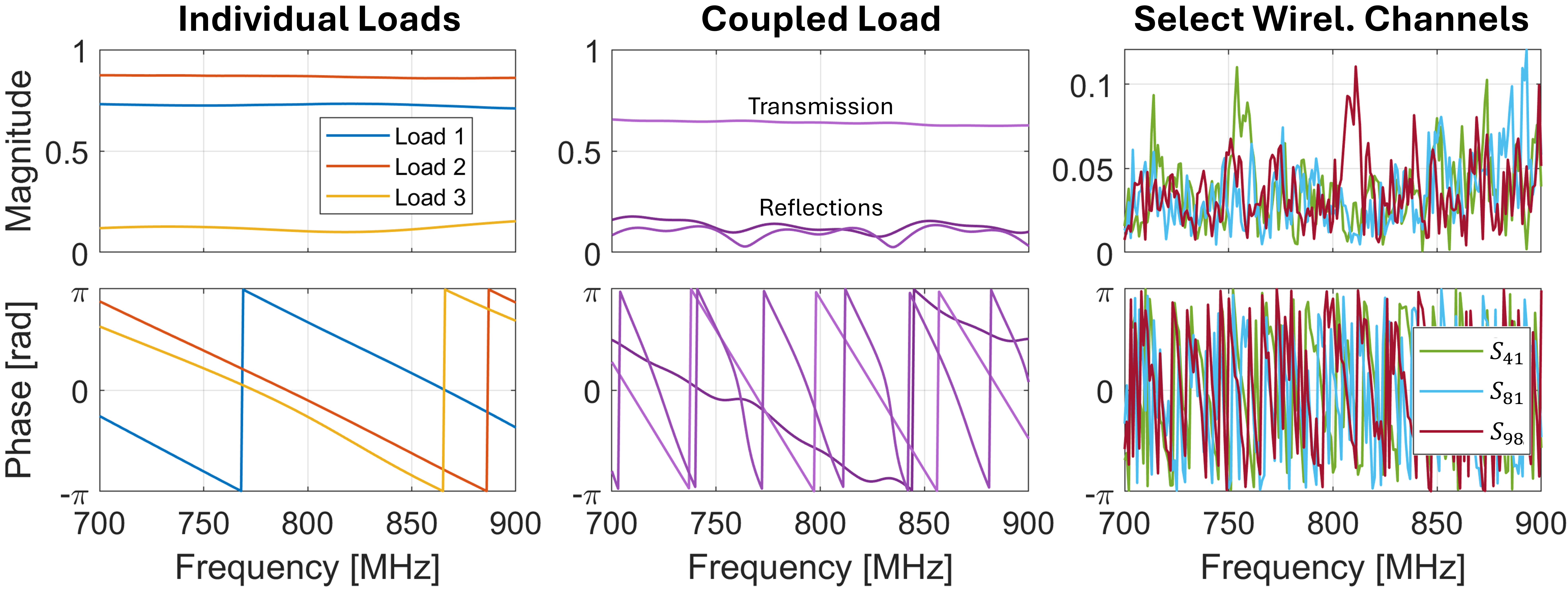}
\caption{Measured scattering characteristics of the tunable load network (left and middle column) and the rich-scattering radio environment (right column, showing three selected wireless channels from the blocks $\mathcal{RT}$, $\mathcal{ST}$ and $\mathcal{SS}$).}
\label{fig2}
\end{figure}

Thanks to our modular design, we directly measure the characteristics of our tunable load network's individual and coupled loads with one-port and two-port VNA measurements, respectively. The measured scattering characteristics are shown in the left and middle columns of Fig.~\ref{fig2}, displaying only weak frequency selectivity. Based on these known characteristics, we can straightforwardly determine $\mathbf{S}_\mathrm{L}$ for any realizable configuration of the tunable load network.

Besides knowing $\mathbf{S}_\mathrm{L}$, the MNT model in (\ref{eq1}) also requires knowledge of certain blocks of $\mathbf{S}$. Thanks to our modular design, all 15 ports defining $\mathbf{S}$ can be accessed in principle, but no 15-port VNA was available. In fact, VNAs commonly have only 2 or 4 ports; despite our access to an 8-port VNA (two cascaded Keysight P5024B 4-port VNAs), we could hence not easily measure $\mathbf{S}$ with a single VNA measurement. Instead of realizing a time-consuming and error-prone measurement of $\mathbf{S}$ by manually reconnecting different sets of antenna ports to the VNA while terminating the others with matched loads~\cite{tippet1982rigorous}, we leveraged the aforementioned Virtual VNA technique~\cite{del2024virtual2p0,tapie2025scalable} that does not require any manual reconnections. Technical details about the Virtual VNA measurement of $\mathbf{S}$ are provided in the Appendix. The frequency dependence of three selected, representative entries of $\mathbf{S}$ are shown in the right column of Fig.~\ref{fig2}, displaying strong frequency selectivity of the wireless channels in the rich-scattering environment.

\section{Performance Evaluation}
\label{sec_PerfEval}

Having characterized both the static radio environment (in terms of $\mathbf{S}$) and the tunable load network (in terms of the possible realizations of $\mathbf{S}_\mathrm{L}$) in the previous section, we can compute in software the end-to-end wireless channel matrix for any realizable BD-RIS configuration in our experiment using (\ref{eq1}). This enables the extensive performance evaluations reported in this section.

\subsection{Key performance indicators}
\label{subsec_KPI}

For clarity, we limit our performance analysis to one single-input single-output (SISO) key performance indicator (KPI) and three multiple-input multiple-output (MIMO) KPIs. We observed very similar trends also for diverse other KPIs. 

For the SISO case, we choose one transmitter and one receiver, such that $\mathbf{H}$ collapses to a scalar $h$.

\textit{SISO KPI:} 
Our SISO KPI is the channel gain $|h|^2$ which we aim to maximize. This is a commonly considered KPI in many theoretical papers on BD-RIS~\cite{li2024beyond}.

For the MIMO cases, we choose two transmitters and two receivers such that $\mathbf{H}\in\mathbb{C}^{2\times 2}$. We assume full channel state information at the transmitters in all cases. In the following, $h_{ji}$ denotes the end-to-end channel from the $i$th transmitter to the $j$th receiver. We define three distinct MIMO KPIs:

\textit{MIMO KPI 1:} We assume that TX1 communicates with RX1 and that TX2 communicates with RX2, such that the signals from TX1 to RX2 and  from TX2 to RX1 constitute undesired interferences. The rate achieved by TX1 is hence $R_1=\mathrm{log}_2\left( 1 + P_\mathrm{T}|h_{11}|^2 / (P_\mathrm{T}|h_{12}|^2 + \sigma^2) \right)$, where we assume that both TX1 and TX2 transmit with power $P_\mathrm{T}$ and $\sigma$ quantifies the noise strength. This scenario is representative of a SISO interference channel. The rate achieved by TX2 is defined analogously. Our KPI is $R=R_1+R_2$, assuming $P_\mathrm{T}/\sigma^2=100$~dB.

\textit{MIMO KPI 2:} We aim to maximize the capacity at low signal-to-noise ratio (SNR). In this scenario, the capacity $C$ is maximized by dominant eigenmode transmission and given by $C=\mathrm{log}_2\left( 1+ P_\mathrm{T}\|\mathbf{H}\|^2/\sigma^2 \right)$. Maximizing $C$ is thus equivalent to maximizing $\|\mathbf{H}\|^2$. Hence, our KPI is $\|\mathbf{H}\|^2$.

\textit{MIMO KPI 3:}  We aim to maximize the capacity at high SNR. In this scenario, the capacity $C$ is maximized by multi-eigenmode transmission with uniform power allocation and given by $C=\mathrm{log}_2\left( \mathrm{det}\left( \mathbf{I}_2+ P_\mathrm{T}\mathbf{H}\mathbf{H}^H/\sigma^2 \right)\right)$. Hence, our KPI is $C$, assuming $P_\mathrm{T}/\sigma^2=100$~dB.

\subsection{Benchmarking}
\label{subsec_benchmarks}

We systematically investigate and benchmark the ability of our  experimental BD-RIS to maximize these four KPIs. We consider the following two benchmarks:

\textit{OC:} All RIS elements are terminated by open-circuit loads (i.e., the tunable load network is absent) such that $\mathbf{S}_\mathrm{L}=\mathbf{I}_{N_\mathrm{S}}$.

\textit{D-RIS:} We consider the same experimental setup but assume that the coupled loads are not available.

In addition, to better understand the influence of hardware constraints, we consider the following alleviated or tightened hardware constraints, both for BD-RIS and D-RIS:

\textit{Ideal}: As in most theoretical studies on BD-RIS~\cite{li2024beyond}, we assume a fully-connected BD-RIS such that all entries  of $\mathbf{S}_\mathrm{L}$ are continuously tunable, subject only to unitarity and reciprocity constraints. For D-RIS, only diagonal entries  of $\mathbf{S}_\mathrm{L}$ are tunable subject to the same constraints.

\textit{1-bit:} We assume that only two of the three individual loads of our tunable load network are available, resulting in 1-bit tunable individual loads. We consider all three choices regarding which two individual loads are retained, denoting them accordingly by `12', `13', and `23'. 

Furthermore, we consider an interleaved BD-RIS (\textit{IBD-RIS}) for which the RIS elements indexed 1, 2, 3, 4, 5, 6, 7, and 8 are connected to the PCB ports indexed 1, 5, 2, 6, 3, 7, 4, and 8, respectively. A recent theoretical study suggests that IBD-RIS can substantially outperform BD-RIS~\cite{nerini2024static}.

\begin{figure*}
\centering
\includegraphics[width=\textwidth]{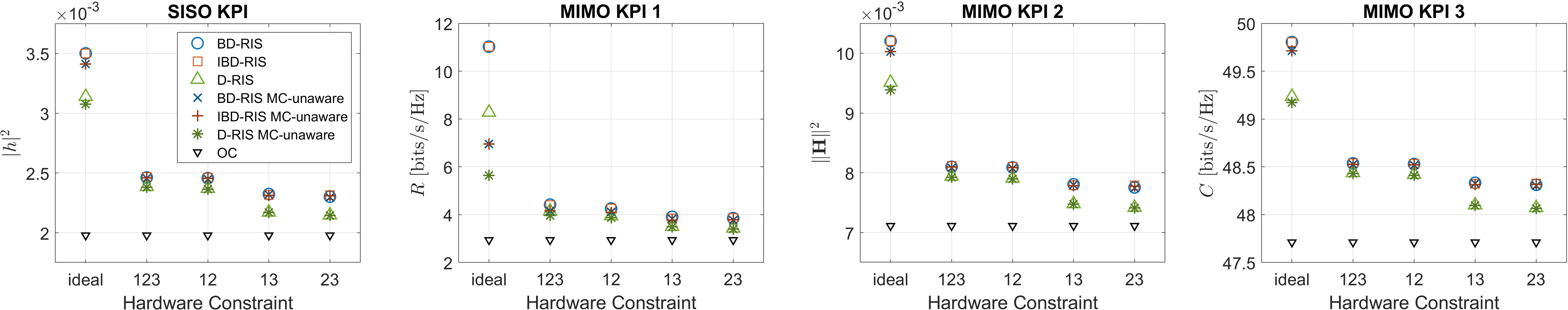}
\caption{Performance evaluation regarding the KPIs defined in Sec.~\ref{subsec_KPI} for our BD-RIS prototype and the benchmarks defined in Sec.~\ref{subsec_benchmarks}. Each displayed point is the average over the KPIs achieved for 201 frequency points and all possible choices of TX and RX among the available ones (see Sec.~\ref{subsec_optimization}). }
\label{fig3}
\end{figure*}

\subsection{Optimization}
\label{subsec_optimization}

For each frequency point and each possible choice of TX and RX among the available ones, we evaluate the four KPIs for each possible configuration of the tunable load network. The small-scale nature of our BD-RIS prototype enables this exhaustive search strategy which ensures that we find the globally optimal configuration in each case. 
In total, there are $\sum_{m=0}^{\lfloor N_S/2\rfloor}
\binom{N_S - m}{m}3^{N_S - 2m}
=12970$ configurations of our BD-RIS. The subsets thereof corresponding to a 1-bit constraint on the individual load programmability contain  $\sum_{m=0}^{\lfloor N_S/2\rfloor}
\binom{N_S - m}{m}2^{N_S - 2m}
=985$ configurations. For the D-RIS benchmarks, the number of possible configurations is $3^8=6561$ with three individual loads and $2^8=256$ under a 1-bit constraint on the individual load programmability.  

For the \textit{ideal} hardware benchmark in which $\mathbf{S}_\mathrm{L}$ is only subject to reciprocity and unitary constraints, we find an optimized configuration for each case using a quasi-Newton algorithm for unconstrained multi-variable optimization. We conveniently leverage the impedance representation $\mathbf{S}_\mathrm{L} =   (\mathbf{Z}_\mathrm{L}  + Z_0\mathbf{I}_{N_\mathrm{S}})^{-1} (\mathbf{Z}_\mathrm{L}  - Z_0\mathbf{I}_{N_\mathrm{S}}) $ and impose that $\mathbf{Z}_\mathrm{L} $ is a purely imaginary and symmetric matrix to satisfy the unitarity and reciprocity constraints, respectively. Then, we optimize the upper triangular part of $\mathbf{Z}_\mathrm{L}$~\cite{shen2021modeling}. 

We also explore how an unawareness of MC during optimization influences the KPIs. To that end, we use (\ref{eq2}) to optimize the RIS configuration but we evaluate the final performance using (\ref{eq1}). Theoretical studies suggest that significant performance degradations can result from ignoring the MC between RIS elements during optimization~\cite{qian2021mutual,li2024beyond}.

\subsection{Results}

The results of our extensive performance evaluations are summarized in Fig.~\ref{fig3}. Each displayed point is the average over the KPI evaluated independently for 201 frequency points and all possible choices of TX and RX among the available ones. TX and RX antennas that are not used in a given scenario are terminated with matched loads.

The results for all four KPIs in Fig.~\ref{fig3} confirm that having access to reconfigurable coupled loads improves the performance. Specifically, considering the `123' hardware constraint which makes full use of our experimentally realized tunable load network, the four KPIs are improved relative to the \textit{OC} benchmark by 24.5\%, 50.3\%, 13.8\%, and 1.7\%, respectively. A strong dependence of the performance improvement on the KPI is apparent. Especially for the first two KPIs the performance improvements are notable despite the small-scale nature of our prototype with only eight RIS elements. Meanwhile, no notable difference between the performances of BD-RIS and IBD-RIS are apparent in our results. 
If the coupled loads are assumed to be unavailable, the performance improvements relative to the \textit{OC} benchmark drop to 20.6\%, 41.0\%, 11.6\%, and 1.5\%, respectively, evidencing the benefit of the coupled loads. The coupled loads are particularly beneficial for the first two KPIs that appear to depend more strongly on the RIS configuration than the last two KPIs. 

Our benchmarks of tighter hardware constraints in which only two of the three individual loads are assumed to be available reveal, both for BD-RIS and D-RIS, that the choice of the available individual loads matters. Specifically, assuming that the third absorptive individual load is unavailable barely affects the KPIs whereas assuming that one of the two reflective individual loads is unavailable does result in notable performance deteriorations. This finding emphasizes that not only the number of available individual loads but also their exact scattering properties matter substantially. Under a constraint to two individual loads of which one is absorptive, the benefit of having reconfigurable coupled loads is seen to be larger for all four KPIs than if all three of our individual loads are available.

Our benchmarks with alleviated hardware constraints yield performance improvements of 76.9\%, 274.7\%, 43.4\%, and 4.4\% for ideal fully-connected BD-RIS and 58.6\%, 181.4\%, 33.7\%, and 3.2\% for ideal  conventional RIS. Clearly, alleviating the hardware constraints by considering those commonly assumed in theoretical studies not only drastically increases the performance gains but also the gap between BD-RIS and D-RIS. 

Finally, we observe that unawareness of MC during optimization barely affects the KPIs in our experimental setup. The same conclusion applies to benchmarks with tightened hardware constraints whereas the influence of MC-unawareness is more visible under alleviated hardware constraints. In particular for MIMO KPI 1, ignoring MC during optimization is found to almost half the achievable performance improvements under an assumption of ideal hardware. Interestingly, it is precisely MIMO KPI 1 that is also the most sensitive to the RIS configuration.

\section{Conclusion}
\label{sec_Conclusion}

To summarize, we have presented the first reflective BD-RIS hardware prototype and performed extensive performance evaluations and benchmarking in a rich-scattering radio environment. Our results confirm the benefit of reconfigurable connections between RIS elements. We observed significant variations in performance gains across different KPIs. Moreover, we observed that realistic hardware constraints substantially influence, \textit{first}, the achievable performance, \textit{second}, the relative benefits of BD-RIS over conventional D-RIS, and, \textit{third}, the influence of MC-unawareness during optimization. 

Looking forward, our work hence encourages further investigations of how realistic hardware constraints may affect the conclusions of recent theoretical studies that assumed ideal hardware. Moreover, important trade-offs related to hardware constraints remain unexplored to date and can be examined in future work, such as the choice arising in our prototype design between having five optimized individual loads as opposed to three optimized individual loads and two coupled loads.

\appendix

In this Appendix, we describe the Virtual VNA measurement of $\mathbf{S}$.
We start by connecting our 8-port VNA to the 7 antenna ports as well as the auxiliary side port next to the first PCB port (auxiliary side ports are indicated by gray circles in Fig.~\ref{fig1}). We configure the first switch to connect the first PCB port to this auxiliary side port. Then, we measure the accessible 8-port scattering matrix for 900 random configurations of the other seven switches. We ensure that if in any given configuration one switch is connected to a coupled load, then so is the other switch associated with that coupled load. The coupled load between the first and second PCB ports is hence never used. Based on these 900 measurements, we leverage a gradient descent procedure similar to those described in~\cite{del2024virtual2p0,tapie2025scalable} to estimate $\mathbf{S}$. The only remaining ambiguity concerns a block-wise sign ambiguity of the $\mathcal{AS}$ and $\mathcal{SA}$ blocks of $\mathbf{S}$, where $\mathcal{A}=\mathcal{T}\cup\mathcal{R}$. To determine the ambiguous sign, we connect the first and second switch to the so far unused coupled load and conduct a measurement of the accessible 7-port scattering matrix. The sign choice that better explains the measurement is retained. Further procedural details can be found in~\cite{del2024virtual2p0,tapie2025scalable}.

\bibliographystyle{IEEEtran}

\providecommand{\noopsort}[1]{}\providecommand{\singleletter}[1]{#1}%

\end{document}